\documentclass[twocolumn,showkeys,aps,prb,showpacs]{revtex4-1}
\usepackage{graphicx}
\usepackage[CJKbookmarks,dvipdfm,colorlinks,linkcolor=blue,citecolor=blue]{hyperref}

\begin{document}

\title{Thermoelectric properties of  $\beta$-As, Sb and Bi monolayers}

\author{Dong-Chen Zhang, Ai-Xia Zhang and San-Dong Guo}
\affiliation{School of Physics, China University of Mining and
Technology, Xuzhou 221116, Jiangsu, China}
\begin{abstract}
Monolayer semiconductors of group-VA elements (As, Sb, Bi) with graphenelike buckled structure offer a potential to achieve nanoscale electronic,  optoelectronic and  thermoelectric devices. Motivated  by recently-fabricated Sb monolayer (antimonene), we systematically investigate the thermoelectric properties of $\beta$-As, Sb and Bi monolayers by  combining the first-principles calculations and semiclassical Boltzmann transport theory.  The generalized gradient approximation (GGA) plus spin-orbit coupling (SOC)  is adopted for the electron part,  and GGA is employed  for the phonon part.
It is found that SOC has important influences on their electronic structures, especially for Bi monolayer, which can induce
observable SOC effects on  electronic   transport coefficients. More specifically, SOC not only has  detrimental influences on  electronic   transport coefficients, but also produces  enhanced effects.
The calculated lattice thermal conductivity  decreases gradually   from As to Bi monolayer, and the corresponding room-temperature sheet thermal conductance is 161.10 $\mathrm{W  K^{-1}}$,   46.62 $\mathrm{W  K^{-1}}$ and 16.02 $\mathrm{W  K^{-1}}$,  which can be  converted  into common lattice thermal conductivity  by dividing by the thickness of 2D material.
The sheet thermal conductance of Bi monolayer is lower than one of other 2D materials, such as  semiconducting transition-metal dichalcogenide monolayers  and orthorhombic group IV-VI monolayers. A series of scattering time is employed to  estimate the thermoelectric figure of merit $ZT$. It is found that  the n-type doping has more excellent thermoelectric properties than p-type doping for As and Bi monolayer, while the comparative $ZT$ between n- and p-type doping  is observed in Bi monolayer.
These results can  stimulate further experimental works to open the new field for thermoelectric devices based on monolayer of group-VA elements.

\end{abstract}


\maketitle

\section{Introduction}
Due to high mobility, heat conductance, and mechanical strength, graphene with a planar honeycomb  structure is one of the most famous  materials, but it lacks an intrinsic band gap, hindering its applications in electronics and optoelectronics\cite{q1}.
The successful exfoliation of graphene intrigues  a large number of search for further two-dimensional (2D) materials like  silicene\cite{q2}, germanene\cite{q3}, phosphorene\cite{q4} and transition-metal dichalcogenides\cite{q5}, etc. Recently, monolayer semiconductors of group-VA elements (As, Sb, Bi) are predicted to be of good
stability with intrinsic gap by the first-principle calculations, and the  graphenelike buckled structure ($\beta$-phase) has the best stability\cite{q6}. The Sb monolayer (antimonene) of them  has been  successfully exfoliated through micromechanical technology\cite{q8}, or has been synthesized on various substrates via van der Waals epitaxy growth\cite{q9}. Experimentally, it has been proved that Sb monolayer is highly stable in ambient conditions\cite{q8}, and the high stability has been observed after aging in air for 30 days\cite{q9}. Raman spectroscopy and transmission electron microscopy
show that the Sb monolayer has graphenelike buckled structure\cite{q8,q9}. It is  urgent and interesting  to study  thermoelectric properties of Sb monolayer in detail, to determine whether it is a potential thermoelectric material.

\begin{figure}
  \includegraphics[width=5.0cm]{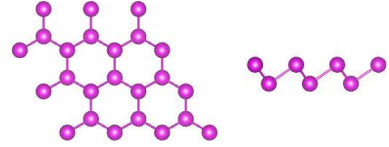}
  \caption{The  crystal structures of $\beta$-As, Sb and Bi monolayers: the top view (Left) and the side view (Right). }\label{struc}
\end{figure}
\begin{figure*}[htp]
  \includegraphics[width=12cm]{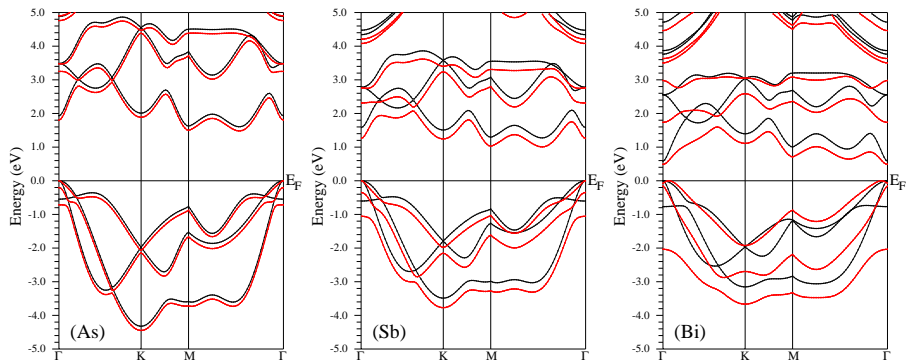}
  \caption{The energy band structures of $\beta$-As, Sb and Bi monolayers within GGA (Black lines) and GGA+SOC (Red lines).}\label{band}
\end{figure*}

Thermoelectric  materials,  which can achieve direct hot-electricity conversion without moving parts, are of great interest in
energy-related issues\cite{s1,s2}. The efficiency of thermoelectric conversion  can be measured by dimensionless  figure of merit, $ZT=S^2\sigma T/(\kappa_e+\kappa_L)$,   in which S, $\sigma$, T, $\kappa_e$ and $\kappa_L$ are the Seebeck coefficient, electrical conductivity, working temperature, the electronic and lattice thermal conductivities, respectively.
It is believed that nanostructured materials, especially for 2D materials, have potential application in highly efficient thermoelectric devices\cite{q10,q11,q12,q13,q14}.  The thermoelectric properties of  semiconducting transition-metal dichalcogenide monolayers and orthorhombic group IV-VI monolayers have been widely investigated \cite{t1,t2,t3,t4,t5,t6}.
Recently, our calculated results show that SOC can produce important effects on electronic transport coefficients of various 2D materials\cite{t4,t6,t7,t8}, so it is necessary for thermoelectric properties of 2D materials to include SOC.
For group-VA monolayers, most of the research are focused on their lattice  thermal conductivities, and their  electronic transports  are little investigated. The  lattice  thermal conductivities of As and Sb monolayers with both graphenelike buckled and black phosphorenelike puckered honeycomb structures ($\alpha$-phase) have been performed in theory\cite{l1,l2,l3,l4}. A highly anisotropic thermal conductivity  along the zigzag and armchair directions is predicted for $\alpha$-As and Sb monolayers\cite{l2,l3}.  It has also been proved that chemical functionalization  can make $\kappa_L$ of Sb monolayer decrease greatly\cite{l4}. The  thermoelectric performance of $\alpha$-As monolayer  has been studied with Green's function based transport techniques,  including both electron and phonon parts\cite{l5}. It  is noteworthy that  a remarkable discrepancy between ref.\cite{l1} and  ref.\cite{l2} for $\kappa_L$ of $\beta$-Sb monolayer can be found.  It may be because the different thickness of Sb monolayer is used, and the related discussion for calculations of transport coefficients of 2D materials can be found in the next section (computational detail).

Here, we  investigate  systematically the thermoelectric properties  of $\beta$-As, Sb and Bi monolayers from ab initio
calculations in combination with Boltzmann transport equation formalism, including both electron and phonon parts. The SOC is
considered for electron part, which is very important for electronic structures and transport coefficients of $\beta$-As, Sb and Bi monolayers, especially for Bi monolayer. The lattice thermal conductivities using the same thickness and sheet thermal conductances of $\beta$-As, Sb and Bi monolayers are calculated, and they gradually decreases from As to Bi monolayer.
It is found that the sheet thermal conductance of Bi monolayer is lower than one of familiar 2D materials. Finally, possible $ZT$
 is estimated by a series of  empirical scattering time, which suggests that $\beta$-As, Sb and Bi monolayers may be potential
 2D thermoelectric materials.

The rest of the paper is organized as follows. In the next section, we shall
describe computational details about the first-principle and transport coefficients calculations. In the third section, we shall present the electronic structures and  thermoelectric properties of $\beta$-As, Sb and Bi monolayers. Finally, we shall give our discussions and conclusion in the fourth section.
\begin{table}[!htb]
\centering \caption{The  lattice constants $a$ ($\mathrm{{\AA}}$) and  buckling parameter $h$  ($\mathrm{{\AA}}$)\cite{q6}; the calculated gap values  using GGA $G$ (eV) and GGA+SOC $G_{so}$ (eV); $G$-$G_{so}$ (eV);  spin-orbit splitting $\Gamma_{\Delta_{so}}$ and  $\mathrm{K_{\Delta_{so}}}$ (eV)  at the high symmetry point $\Gamma$ and $\mathrm{K}$ of valence bands. }\label{tab}
  \begin{tabular*}{0.48\textwidth}{@{\extracolsep{\fill}}cccccccc}
  \hline\hline
Name& $a$ & $h$ & $G$& $G_{so}$&$G$-$G_{so}$ &$\Gamma_{\Delta_{so}}$& $K_{\Delta_{so}}$\\\hline\hline
As&3.61&1.40&1.61&1.48&0.13  &0.21&0.07\\\hline
Sb&4.12&1.65&1.27&1.01& 0.26 &0.35&0.17\\\hline
Bi&4.34&1.73&0.58&0.49& 0.09 &0.08&0.77\\\hline\hline
\end{tabular*}
\end{table}
\begin{figure*}[htp]
  \includegraphics[width=16cm]{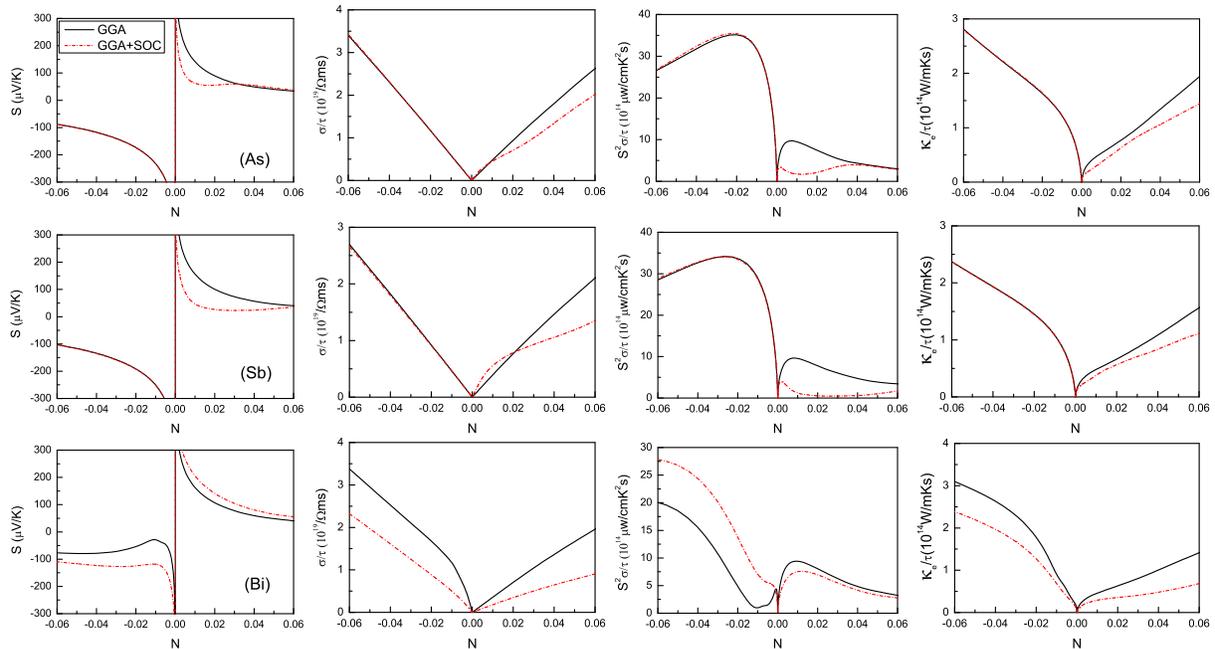}
  \caption{(Color online) At room temperature (300 K),  transport coefficients of $\beta$-As, Sb and Bi monolayers as a function of doping level (N) within GGA and GGA+SOC:  Seebeck coefficient S,  electrical conductivity with respect to scattering time  $\mathrm{\sigma/\tau}$,  power factor with respect to scattering time $\mathrm{S^2\sigma/\tau}$  and electronic thermal conductivity with respect to scattering time $\mathrm{\kappa_e/\tau}$. }\label{s0}
\end{figure*}

\section{Computational detail}
 We use a full-potential linearized augmented-plane-waves method
within the density functional theory (DFT) \cite{1} to carry out electronic structures of  $\beta$-As, Sb and Bi monolayers, as implemented in the package WIEN2k \cite{2}. The  free  atomic position parameters  are optimized using GGA of Perdew, Burke and  Ernzerhof  (GGA-PBE)\cite{pbe} with a force standard of 2 mRy/a.u..
The SOC is included self-consistently \cite{10,11,12,so}, which gives rise to important influences on electronic transport coefficients. The convergence results are determined
by using  5000 k-points in the
first Brillouin zone (BZ) for the self-consistent calculation, making harmonic expansion up to $\mathrm{l_{max} =10}$ in each of the atomic spheres, and setting $\mathrm{R_{mt}*k_{max} = 8}$ for the plane-wave cut-off. The self-consistent calculations are
considered to be converged when the integration of the absolute
charge-density difference between the input and output electron
density is less than $0.0001|e|$ per formula unit, where $e$ is
the electron charge.

Based on the results of electronic
structure, transport coefficients for electron part
are calculated through solving Boltzmann
transport equations within the constant
scattering time approximation (CSTA),  as implemented in
BoltzTrap\cite{b}, which shows reliable results in many classic thermoelectric
materials\cite{b1-1,b2,b3}. To
obtain accurate transport coefficients, we set the parameter LPFAC for 20, and use at least 2408 k-points  in the  irreducible BZ for the energy band calculation. The  lattice thermal conductivities are performed
by using Phono3py+VASP codes\cite{pv1,pv2,pv3,pv4}. The second order harmonic and third
order anharmonic interatomic force constants  are calculated by using a  5 $\times$ 5 $\times$ 1   supercell  and a  4 $\times$ 4 $\times$ 1 supercell, respectively. To compute lattice thermal conductivities, the
reciprocal spaces of the primitive cells  are sampled using the 50 $\times$ 50 $\times$ 2  meshes.

For 2D material, the calculated  electrical conductivity, electronic  and lattice  thermal conductivities  depend on the length of unit cell used in the calculations along z direction\cite{2dl}, which should be normalized by multiplying Lz/d, where Lz is the length of unit cell along z direction  and d is the thickness of 2D material, but the d  is not well defined.  However, The dimensionless figure of merit $ZT$ is independent of  the length of unit cell used in the calculations along z direction. In this work, the length of unit cell used in our calculations along z direction is used as the thickness of $\beta$-As, Sb and Bi monolayers, and the corresponding d is 18 $\mathrm{{\AA}}$. The thermal sheet conductance can be used as a fair comparison between various 2D monolayers, which can be attained by $\kappa$ $\times$ d.

\begin{figure*}[htp]
  \includegraphics[width=12cm]{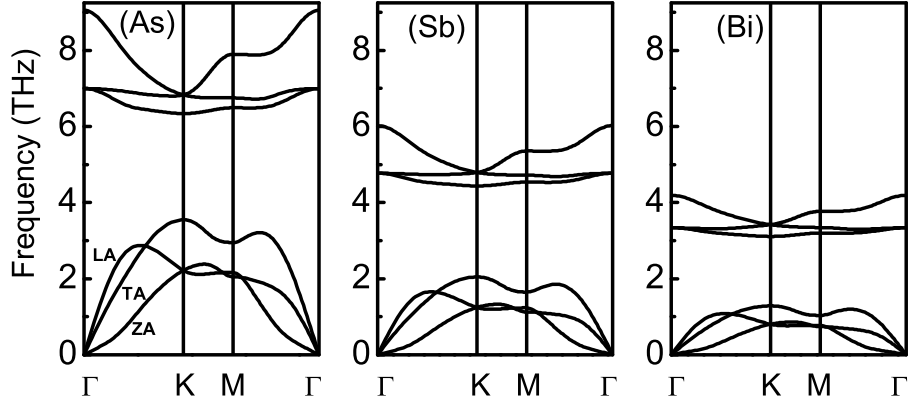}
  \caption{Phonon band structure of $\beta$-As, Sb and Bi monolayers using GGA-PBE.}\label{ph}
\end{figure*}

\begin{figure*}[htp]
  \includegraphics[width=12cm]{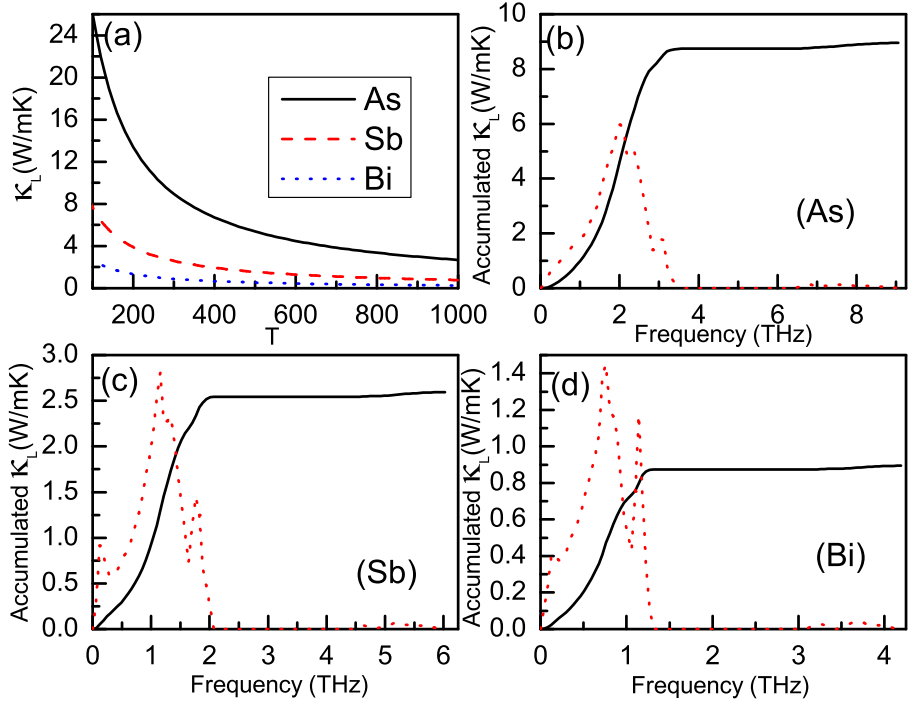}
  \caption{(a) The lattice thermal conductivities of  $\beta$-As, Sb and Bi monolayers  as a function of temperature using GGA-PBE. (b), (c) and (d) The accumulated lattice thermal conductivities of  $\beta$-As, Sb and Bi, and the derivatives.}\label{kl}
\end{figure*}

\section{MAIN CALCULATED RESULTS AND ANALYSIS}

The $\beta$-phase of As, Sb and Bi monolayers is a graphenelike buckled honeycomb structure with space group of $P\bar{3}m1$ (No. 164), and  the schematic crystal structure is shown in \autoref{struc}. The $\beta$-phase is different from $\alpha$-phase
with puckered honeycomb structure.  The unit cell of $\beta$-phase contains two atoms with each
atom  connected to three  atoms of another plane, while $\alpha$-phase contains four atoms
with each atom connected to two atoms of the same plane
and one atom of another plane.
In this work, the optimized lattice constants of $\beta$-phase are used\cite{q6}, which are summarized in \autoref{tab}. It is found that the buckling parameter $h$ from \autoref{tab}, defined as the vertical distance separating the two atomic planes,
gradually increases from As to Bi monolayer.
The unit cell  of $\beta$-As, Sb and Bi monolayers   is constructed with the vacuum region of larger than 16 $\mathrm{{\AA}}$ to avoid spurious interaction. As is well known, the SOC has very important effects on electronic structures and
 electronic   transport coefficients of materials containing heavy element\cite{t4,t6,t7,t8}, such as Bi. Firstly, the electronic structures of $\beta$-As, Sb and Bi monolayers are investigated using GGA and GGA+SOC, and the energy band structures are plotted in \autoref{band} using GGA and GGA+SOC. The  As and Sb monolayers  are indirect band gap semiconductors with the  valence band maximum  (VBM) at the $\Gamma$ point and conduction band minimum (CBM)  at the M point using both GGA and GGA+SOC. The Bi monolayer is a direct band gap semiconductor with VBM and  CBM at the $\Gamma$ point using GGA, while it is a indirect band gap semiconductors with VBM between  the $\Gamma$ and M points and CBM  at the $\Gamma$ point using GGA+SOC.
 The  GGA and GGA+SOC gaps,  and the differences between them are listed in \autoref{tab}. Both GGA and GGA+SOC gaps decrease from As to Bi monolayer, and the difference of Bi monolayer is less than ones of As and Sb monolayers, which is due to the change of CBM  from M to $\Gamma$ point. The three monolayers  have  some  conduction band extrema (CBE) around the Fermi level, which is favorable for n-type Seebeck coefficient. According to SOC-induced changes of outlines of energy bands, the strength of SOC  increases from As to Bi monolayer, which is  consistent with their  respective atomic mass.  The representative spin-orbit splitting values    at the high symmetry points $\Gamma$ and $\mathrm{K}$ of valence bands  are summarized in \autoref{tab}. It is found that the spin-orbit splitting value at  $\mathrm{K}$  point can represent the strength of SOC.
\begin{figure*}[htp]
  \includegraphics[width=15cm]{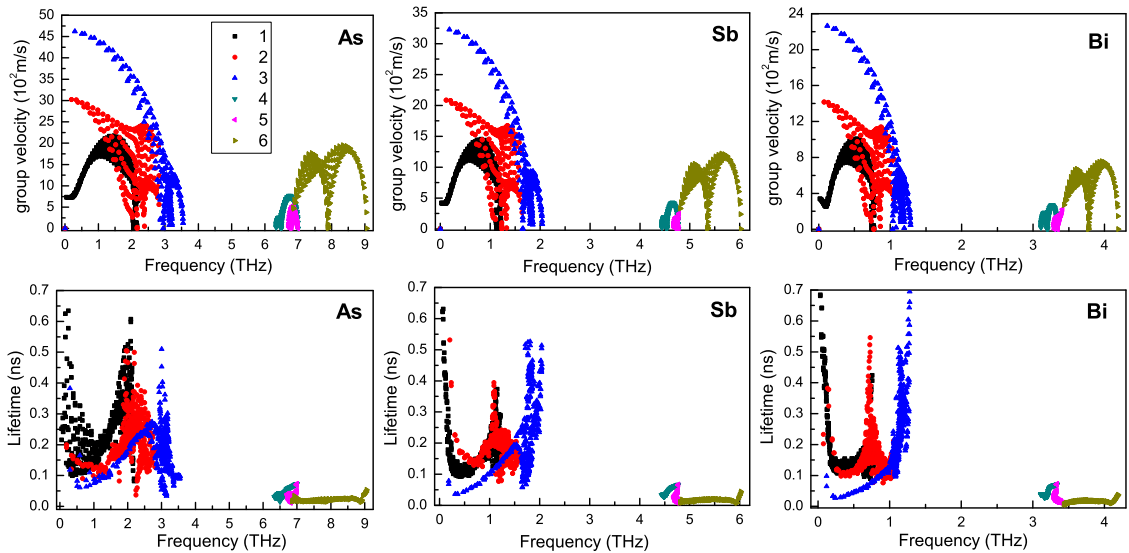}
  \caption{(Color online) Calculated phonon group velocities  and phonon lifetimes of $\beta$-As, Sb and Bi monolayers  using GGA-PBE in the irreducible Brillouin zone. 1 represents ZA branch, 2 for TA branch, 3 for LA branch and 4, 5, 6 for optical branches.}\label{szt}
\end{figure*}

 The SOC can produce very important effects on electronic transport coefficients in many 2D materials, such as semiconducting transition-metal dichalcogenide  and orthorhombic group IV-VI monolayers\cite{t4,t6,t7,t8}. Firstly, the semi-classic transport coefficients are calculated  within CSTA Boltzmann theory using GGA and GGA+SOC, and  the doping effects are simulated  by  simply shifting the Fermi level into conduction or valence bands within rigid band approach, which is valid with the doping level being low\cite{tt9,tt10,tt11}. The Seebeck coefficient S,  electrical conductivity with respect to scattering time  $\mathrm{\sigma/\tau}$,  power factor with respect to scattering time $\mathrm{S^2\sigma/\tau}$  and electronic thermal conductivity with respect to scattering time $\mathrm{\kappa_e/\tau}$ as  a function of doping level (N) at  room temperature  are plotted in \autoref{s0} using GGA and GGA+SOC.

 For As and Sb monolayers, it is found that SOC has a detrimental effect on Seebeck coefficient in p-type doping, while has  a  negligible influence  in n-type doping. The detrimental effect can be explained by SOC-induced splitting at $\Gamma$ point, which reduces the degeneracy of energy bands. However, the enhanced effect on Seebeck coefficient (absolute value)  can be observed  in both n- and p-type doping for Bi monolayer, which can be understood by bands convergence\cite{s1}.  At presence of SOC, the CBE along $\Gamma$-M and CBM  approach each other, which leads to improved n-type Seebeck coefficient.   When SOC is considered,
  the VBM changes from $\Gamma$ point to one point along $\Gamma$-M, and another  valence  band extrema (VBE) appears at one point along $\Gamma$-K. The VBE along $\Gamma$-K and  VBM   are  very  close,  and the energy difference only is  0.0005 eV, which gives rise to enhanced p-type  Seebeck coefficient.
\begin{table}[!htb]
\centering \caption{The maximal acoustic vibration frequency $MAVF$ (THz),  width of optical branches $WO$ (THz),
   phonon band gap $PBG$ (THz),  thermal sheet conductance $TSC$ ($\mathrm{W K^{-1}}$),   acoustic branch  contribution to lattice thermal conductivity $ACL$ (\%) and  corresponding frequency of peak value of derivatives $FPD$ (THz).  }\label{tab2}
  \begin{tabular*}{0.48\textwidth}{@{\extracolsep{\fill}}ccccccc}
  \hline\hline
Name& $MAVF$ & $WO$ &$PBG$ & $TSC$& $ACL$&$FPD$ \\\hline\hline
As&3.55&2.71&2.79&161.10&97.6 &2.01\\\hline
Sb&2.05&1.58&2.39&46.62& 98.1 &1.06\\\hline
Bi&1.29&1.08&1.82&16.02& 98.1 &0.76\\\hline\hline
\end{tabular*}
\end{table}

 In n-type doping,   a neglectful influence on $\mathrm{\sigma/\tau}$ of As and Sb monolayers can be observed at the presence of SOC,  while  a slightly improved effect in low p-type doping and  a detrimental influence in high p-type doping can be achieved. For Bi monolayer, a decreased effect on  $\mathrm{\sigma/\tau}$  caused by SOC   in both n- and p-type doping can be observed. The effect on  $\mathrm{S^2\sigma/\tau}$ of As and Sb monolayers induced by SOC  has the same trend with one on S.
 A enhanced effect  in n-type doping and a decreased  influence in p-type doping on $\mathrm{S^2\sigma/\tau}$ of Bi monolayer produced  by SOC can be fulfilled. It is found  that the effect on $\mathrm{\kappa_e/\tau}$  caused by SOC has almost similar trend with one on $\mathrm{\sigma/\tau}$ , which can be explained by the Wiedemann-Franz law: $\mathrm{\kappa_e}$= $L$$\mathrm{\sigma}$$T$, in which $L$ is the Lorenz number.

Based on the linearized phonon Boltzmann equation within single-mode relaxation time approximation, the lattice  thermal conductivities of $\beta$-As, Sb and Bi monolayers can be attained, which are assumed to be independent of  doping level. Their phonon band structures are shown in \autoref{ph}, which agree well with previous results\cite{q6}. Their unit cell  contains two  atoms, resulting in 3 acoustic
and 3 optical phonon branches.
From As to Bi monolayer,  the whole  branches   move toward low energy, and the phonon dispersion becomes more localized.
The maximal acoustic vibration frequency (MAVF)  is 3.55 THz, 2.05 THz and 1.29 THz from As to Bi monolayer, and a low MAVF  is benefit to low thermal conductivity. The width of optical branches of  $\beta$-As, Sb and Bi monolayers is 2.71 THz, 1.58 THz and 1.08 THz, respectively, and the narrow width suggests little contribution to thermal conductivity. 
 The third optical branch has larger dispersion than  the first two ones, which meas the third one has larger group velocity, leading to obvious contribution to lattice  thermal conductivity.
 The phonon band gap decreases from As (2.79 THz) to Sb (2.39 THz) to Bi (1.82 THz) monolayer. It is found that  the longitudinal
acoustic (LA) and transverse acoustic (TA) branches are linear near the $\Gamma$ point, while the z-direction acoustic (ZA) branch  deviates from linearity near the $\Gamma$ point, which shares the general feature of 2D materials\cite{p1,p2}. The related data are summarized \autoref{tab2}.

\begin{table}[!htb]
\centering \caption{At 300 K, the phonon mode and total lattice thermal conductivities of  $\beta$-As, Sb and Bi monolayers, including acoustic (ZA, TA and LA) and optical (O1, O2 and O3) branches.  Unit: ( $\mathrm{W m^{-1} K^{-1}}$).}\label{tab-new}
  \begin{tabular*}{0.48\textwidth}{@{\extracolsep{\fill}}cccccccc}
  \hline\hline
Name& ZA & TA &LA & O1& O2&O3& $\kappa_L$ \\\hline\hline
As&2.927&3.109&2.702&0.032&0.013 &0.169&8.952\\\hline
Sb&0.630&1.086&0.826&0.009& 0.001 &0.043&2.594\\\hline
Bi&0.247&0.360&0.266&0.004& 0.001 &0.016&0.894\\\hline\hline
\end{tabular*}
\end{table}

\begin{figure*}[htp]
  \includegraphics[width=15cm]{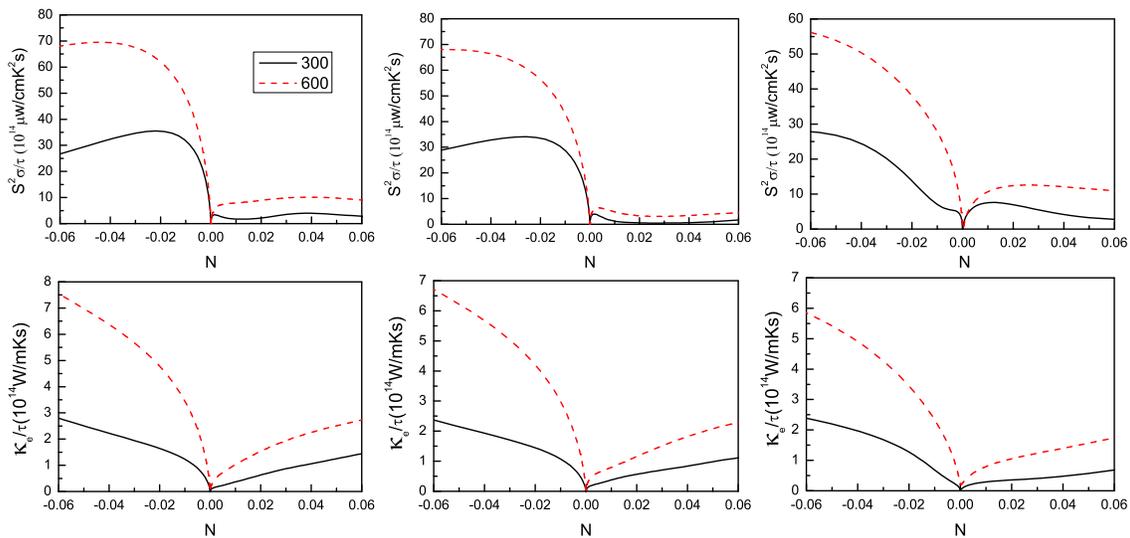}
  \caption{(Color online) At 300 and 600 K, the power factor with respect to scattering time $\mathrm{S^2\sigma/\tau}$  and electronic thermal conductivity with respect to scattering time $\mathrm{\kappa_e/\tau}$ versus doping level (N) using GGA+SOC.}\label{sket}
\end{figure*}

The lattice  thermal conductivities of  $\beta$-As, Sb and Bi monolayers as a function of temperature are shown in \autoref{kl}, and the same thickness d (18 $\mathrm{{\AA}}$) is used. The room-temperature lattice  thermal conductivity of $\beta$-As, Sb and Bi monolayers is 8.95 $\mathrm{W m^{-1} K^{-1}}$, 2.59 $\mathrm{W m^{-1} K^{-1}}$ and 0.89 $\mathrm{W m^{-1} K^{-1}}$, respectively.  To compare the lattice thermal conductivities of various 2D materials, we  convert  all thermal conductivity values into thermal sheet conductance\cite{2dl}, and the corresponding  thermal sheet conductance is  161.10 $\mathrm{W K^{-1}}$,   46.62 $\mathrm{W K^{-1}}$ and 16.02 $\mathrm{W K^{-1}}$, respectively. The thermal sheet conductance of Bi monolayer is lower than one of other 2D materials (semiconducting transition-metal dichalcogenide  and orthorhombic group IV-VI monolayers), and is very close to ones of SnS (18.68 $\mathrm{W K^{-1}}$) and SnSe (17.55 $\mathrm{W K^{-1}}$)\cite{2dl}.
To examine the relative contributions of acoustic  modes to the total lattice thermal conductivity,
the cumulative lattice thermal conductivity  and the derivatives  are also plotted in \autoref{kl} at room temperature.
The acoustic branch of  $\beta$-As, Sb and Bi monolayers provides a contribution of 97.6\%, 98.1\% and 98.1\%, respectively.
 This meets the usual picture that high-frequency optical  phonons  have very little contribution to thermal conductivity.
 The derivatives show that the change of  lattice thermal conductivity  versus frequency has a peak value, and the corresponding frequency of peak value is 2.01 THz, 1.16 THz and 0.76 THz from As to Bi monolayer. These frequencies  and  cross values of ZA and LA branches almost overlap. Some key data are shown in \autoref{tab2}. Furthermore, we examine the relative contributions of six phonon modes to the total lattice
thermal conductivity, and the mode lattice thermal conductivities are shown \autoref{tab-new}. It is found that TA branch  provides the largest  contribution in acoustic branches, and the third optical branch gives the greatest contribution in optical ones.

According to \autoref{szt}, the  group velocity of ZA branch  is smaller than  ones of LA and TA branches, which is due to  nonlinear  dispersion of ZA branch near the $\Gamma$ point. It is clearly seen that group velocities  become small from As to Bi monolayer, which  leads to a decrescent thermal conductivity. The largest  group velocity  for ZA, TA and LA branches near $\Gamma$ point is 0.73  $\mathrm{km s^{-1}}$, 3.03  $\mathrm{km s^{-1}}$ and 4.62  $\mathrm{km s^{-1}}$ for As monolayer, 0.42 
$\mathrm{km s^{-1}}$, 2.08 $\mathrm{km s^{-1}}$ and 3.24 $\mathrm{km s^{-1}}$ for Sb monolayer, 0.34 $\mathrm{km s^{-1}}$, 1.42 $\mathrm{km s^{-1}}$ and 2.26 $\mathrm{km s^{-1}}$ for Bi monolayer. The phonon lifetimes of $\beta$-As, Sb and Bi monolayers  at room temperature are plotted  in \autoref{szt}. As can be seen, most of the phonon lifetimes are decrease from As to Bi monolayer, which can explain decreasing lattice thermal conductivity. It is found that optical  phonon lifetimes are very shorter than acoustic ones, which suggests that optical branches has little contribution to lattice thermal conductivity.

\begin{figure*}[htp]
  \includegraphics[width=15cm]{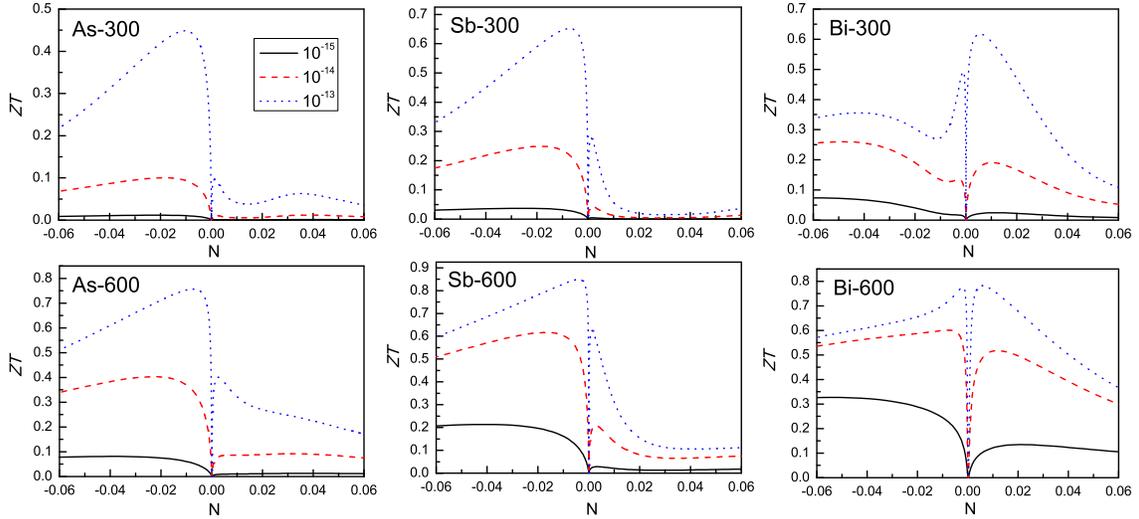}
  \caption{(Color online) At 300 and 600 K, calculated $ZT$ of $\beta$-As, Sb and Bi monolayers  as a function of doping level with the scattering time $\mathrm{\tau}$  being 1 $\times$ $10^{-15}$ s, 1 $\times$ $10^{-14}$ s and 1 $\times$ $10^{-13}$ s.}\label{zt}
\end{figure*}

Based on calculated electron and phonon transport coefficients, the figure of merit $ZT$ can be attained  to estimate the efficiency of thermoelectric conversion.  Firstly,  the power factor with respect to scattering time $\mathrm{S^2\sigma/\tau}$  and electronic thermal conductivity with respect to scattering time $\mathrm{\kappa_e/\tau}$ versus doping level (N) using GGA+SOC at 300 and 600 K are plotted in \autoref{sket}.
However, another unknown quantity is scattering time  $\mathrm{\tau}$, but it is difficulty to  calculate  scattering time  from the first-principle calculations due to the complexity of various carrier scattering mechanisms. Here, 1 $\times$ $10^{-15}$ s, 1 $\times$ $10^{-14}$ s  and 1 $\times$ $10^{-13}$ s are used to attain the power factor  and electronic thermal conductivity.
According to attained power factor, electronic thermal conductivity and lattice  thermal conductivity from \autoref{kl},
the possible $ZT$ of $\beta$-As, Sb and Bi monolayers  as a function of doping level at 300 and 600 K are plotted in \autoref{zt}. It is found that $ZT$ increases with increased scattering time  $\mathrm{\tau}$, which can be explained by the relation: $ZT$=$ZT_e$$\times$$\kappa_e/(\kappa_e+\kappa_L)$, in which  $ZT_e=S^2\sigma T/\kappa_e$ as  an upper limit of $ZT$.  The  $\kappa_e$ can be attained by $\kappa_e/\tau_r$$\times$$\tau$, in which $\tau_r$, $\tau$ are true and empirical scattering time, respectively. As  the scattering time $\mathrm{\tau}$ increases, the $\kappa_e$ increases. That leads to that the $\kappa_e/(\kappa_e+\kappa_L)$ is more close to one, and then  $ZT$ is closer to $ZT_e$.
Calculated results show that the  n-type doping  for As and Sb monolayers has more excellent  $ZT$ than  p-type doping, and that Bi monolayer shows almost equivalent $ZT$ between n- and p-type doping. It is found that a peak $ZT$ in n-type doping is up to 0.40, 0.62 and 0.60 from As to Bi monolayer with classic scattering time $\tau$=$10^{-14}$ s, and the corresponding doping level gradually decreases. These imply that $\beta$-As, Sb and Bi monolayers  may be potential 2D thermoelectric materials by optimizing doping.

\section{Discussions and Conclusion}
Seebeck coefficient can be tuned by removing or enhancing band degeneracy, and further can  affect power factor. For  $\beta$-As, Sb and Bi monolayers, the SOC removes not only band degeneracy (orbital degeneracy), but also enhance one  (valley degeneracy).
The SOC can lead to  reduced effects on the p-type  Seebeck coefficient of As and Sb monolayers by lifting the degenerate of $\Gamma$ point, but can enhance Seebeck coefficient of Bi monolayer due to bands converge caused by spin-orbit splitting.
Similar  SOC effects on  Seebeck coefficient are also found in  semiconducting transition-metal dichalcogenide monolayers $\mathrm{MX_2}$ (M=Zr, Hf, Mo, W and Pt; X=S, Se and Te)\cite{t8}. The  observably enhanced effects on Seebeck coefficient   can be observed  in monolayers $\mathrm{WX_2}$ (X=S, Se and Te), and detrimental effects in $\mathrm{MX_2}$ (M=Zr, Hf, Mo and Pt; X=S, Se and Te). For bulk materials, the detrimental effects on Seebeck coefficient caused by SOC also can be found in $\mathrm{Mg_2X}$ (X = Si, Ge, Sn)\cite{w3,w4} and   half-Heusler ANiB (A = Ti, Hf, Sc, Y; B = Sn, Sb, Bi)\cite{w5}.  Therefore, including SOC  is very important  for electronic   transport coefficients of   $\beta$-As, Sb and Bi monolayers.
\begin{table*}[!htb]
\centering \caption{A ($\mathrm{W m^{-1} K^{-1}}$): Lattice thermal conductivity using the same thickness of the interlayer distance of graphite (3.35 $\mathrm{{\AA}}$). B ($\mathrm{W K^{-1}}$): Thermal sheet conductance.}\label{tab1}
  \begin{tabular*}{0.96\textwidth}{@{\extracolsep{\fill}}cccccccccccc}
  \hline\hline
Name& GeS& GeSe& SnS& SnSe&  $\mathrm{ZrS_2}$& $\mathrm{ZrSe_2}$& $\mathrm{HfS_2}$& $\mathrm{HfSe_2}$&  As&Sb &Bi                                       \\\hline\hline
A&  15.80 &  9.43 &5.58 &5.24 &23.25 & 18.55 &  29.06 &20.71 &   48.09 &  13.74&  4.78                              \\\hline
B&  52.93 &  31.58&18.68&17.55&77.89 &  62.14 &  97.35&  69.38&  161.10&  46.62& 16.02                                            \\\hline\hline
\end{tabular*}
\end{table*}

The electronic structures of 2D materials is quite sensitive to strain, and $\beta$-As, Sb and Bi monolayers have  some  CBE around the Fermi level. So, it is possible to tune their  thermoelectric properties by band engineering.
Strain (pressure) has been proved to be a very effective strategy to improve  thermoelectric properties of 2D (bulk)  materials\cite{t4,t7,w4,e4-1}.
For bulk $\mathrm{Mg_2Sn}$, pressure can induce  accidental degeneracies (orbital degeneracy) of CBM, which can significantly improve power factor\cite{w4}. Strain-enhanced  power factor  can also be found  in monolayer $\mathrm{MoS_2}$\cite{t7}, $\mathrm{PtSe_2}$\cite{t4} and $\mathrm{ZrS_2}$\cite{e4-1} due to  bands converge (valley degeneracy) induced by strain.
It has also been proved that tensile strain can reduce lattice thermal conductivity in many 2D materials, such as $\mathrm{PtSe_2}$\cite{t4}, $\mathrm{PtTe_2}$\cite{gsd} and $\mathrm{ZrS_2}$\cite{e4-1}. It has been predicted that strain can induce   bands converge in As monolayer\cite{b1}. Therefore, strain effects on thermoelectric properties  of $\beta$-As, Sb and Bi monolayers  are  well worth studying. It has been predicted that the perpendicular
electric field can induce an indirect-to-direct gap transition of As monolayer at 4.2 V/nm by the first-principle calculations\cite{b1}, which suggests that the electric field can also tune the electronic structures. Therefore, it is possible to tune electronic transport properties  of $\beta$-As, Sb and Bi monolayers by electric field.

The potential thermoelectric materials should possess low lattice thermal conductivity.
To compare the lattice thermal conductivities of different
monolayer 2D materials,  the same thickness d should be adopted, or the sheet thermal conductance should be used\cite{2dl}. Here,  the same thickness of the interlayer distance of graphite (3.35 $\mathrm{{\AA}}$) is used\cite{2dl}.
 The lattice thermal conductivities  and sheet thermal conductances  of $\beta$-As, Sb and Bi monolayers, some semiconducting transition-metal dichalcogenide monolayers  and orthorhombic group IV-VI monolayers are summarized in \autoref{tab1}.
 The lattice thermal conductivity of Bi monolayer is lower than that of orthorhombic group IV-VI monolayers  and  semiconducting transition-metal dichalcogenide monolayers, which suggests that it may be a potential 2D thermoelectric material compared to other  familiar 2D materials. The lattice thermal conductivity of Sb monolayer has been widely calculated, for example
 15.1 $\mathrm{W m^{-1} K^{-1}}$\cite{l1}, 13.8 $\mathrm{W m^{-1} K^{-1}}$\cite{l2} and 2.3 $\mathrm{W m^{-1} K^{-1}}$\cite{l4}.
This apparent contradiction may be because the different thickness is used to calculate lattice thermal conductivity.

In conclusion, we have carried out a detailed theoretical studies of the thermoelectric properties of $\beta$-As, Sb and Bi monolayers based on ab initio calculations combined with Boltzman transport theory. It is proved that SOC has important
effects on electronic properties of $\beta$-As, Sb and Bi monolayers, which has been ignored in other theoretical calculations\cite{q6,b1}. The sheet thermal conductance is used to compare lattice thermal conductivities of different 2D materials, and the sheet thermal conductance of  Bi monolayer is lower than one of other well-studied 2D materials, being very favorable to realize high thermoelectric  efficiency. Finally, a serials of  hypothetical scattering time is adopted to estimate possible  efficiency of thermoelectric conversion.  Our work suggests that these $\beta$-As, Sb and Bi monolayers  with graphenelike buckled structure may offer a new 2D playground to achieve high-performance  thermoelectric devices.

\begin{acknowledgments}
This work is supported by the National Natural Science Foundation of China (Grant No. 11404391). We are grateful to the Advanced Analysis and Computation Center of CUMT for the award of CPU hours to accomplish this work.
\end{acknowledgments}


\begin{references}
\bibitem{q1}A. H. Castro Neto, F. Guinea, N. M. R. Peres, K. S. Novoselov,
and A. K. Geim, Rev. Mod. Phys. \textbf{81}, 109 (2009).


\bibitem{q2} B. Aufray, A. Kara, S. Vizzini, H. Oughaddou, C. L¨¦andri, B. Ealet and G. Le Lay, Appl.
Phys. Lett.  \textbf{96}, 183102 (2010).


\bibitem{q3}L. Li, Y. Yu, G. J. Ye, Q. Ge, X. Ou, H. Wu, D. Feng, X. H. Chen and Y. Zhang, Nature
Nanotech.  \textbf{9}, 372 (2014).

\bibitem{q4} H. Liu, A. T. Neal, Z. Zhu, Z. Luo, X. Xu, D. Tomnek and P. D. Ye, ACS Nano  \textbf{8},
4033 (2014).

\bibitem{q5}G. Cunningham, M. Lotya, C. S. Cucinotta, S. Sanvito, S. D. Bergin, R. Menzel, M. S. P.
Shaffer and J. N. Coleman, ACS Nano.  \textbf{6}, 3468 (2012).


\bibitem{q6}S. L. Zhang  et al.,  Angew. Chem. \textbf{128}, 1698 (2016).



\bibitem{q8}P. Ares et al., Adv. Mater. \textbf{28}, 6332 (2016).


\bibitem{q9}J. P. Ji  et al.,  Nat. Commun. \textbf{7}, 13352 (2016).


\bibitem{s1} Y. Pei, X. Shi, A. LaLonde, H. Wang, L. Chen and G. J. Snyder, Nature \textbf{473}, 66 (2011).

\bibitem{s2} A. D. LaLonde, Y. Pei, H. Wang and G. J. Snyder, Mater. Today \textbf{14}, 526 (2011).

\bibitem{q10} L. Hicks and M. Dresselhaus, Phys. Rev. B \textbf{47}, 12727 (1993).

\bibitem{q11}L. Hicks, T. Harman and M. Dresselhaus,  Appl. Phys. Lett. \textbf{63}, 3230  (1993).

\bibitem{q12} Y. Xu, Z. Li, W. Duan,  Small  \textbf{10}, 2182 (2014).

\bibitem{q13}J. P.  Heremans, M.  Dresselhaus, L. E.  Bell and D. T. Morelli,  \textbf{8}, 471 (2013).

\bibitem{q14}M. Dresselhaus et al.,  Adv. Mater.   \textbf{19}, 1043 (2007).




\bibitem{t1}S. Kumar and U. Schwingenschl$\ddot{o}$gl, Chem. Mater.  \textbf{27}, 1278 (2015).


\bibitem{t2}J.  Wu  et al.  Nano Lett. \textbf{14}, 2730  (2014).


\bibitem{t3}Z. Jin  et al.  Sci. Rep. \textbf{5}, 18342  (2015).

\bibitem{t4}S. D. Guo, J. Mater. Chem. C  \textbf{4}, 9366 (2016).

\bibitem{t5} F. Q. Wang, S. Zhang, J. Yu and Q. Wang, Nanoscale \textbf{7},
15962 (2015).

\bibitem{t6}S. D. Guo and Y. H. Wang, J. Appl. Phys. \textbf{121}, 034302 (2017).



\bibitem{t7}S. D. Guo, Comp. Mater. Sci. \textbf{123}, 8 (2016).



\bibitem{t8}S. D. Guo and J. L. Wang, Semicond. Sci. Tech.
\textbf{31}, 095011 (2016).




\bibitem{l1}S. D. Wang,   W. H. Wang and    G. J.  Zhao, Phys. Chem. Chem. Phys. \textbf{18}, 31217 (2016)


\bibitem{l2}G. H. Zheng, Y. L. Jia, S. Gao and S. H.  Ke, Phys. Rev. B \textbf{94}, 155448 (2016).


\bibitem{l3}M. Zeraati, S. M. V. Allaei, I. A. Sarsari, M. Pourfath, and D.
Donadio, Phys. Rev. B  \textbf{93}, 085424 (2016).


\bibitem{l4}T.  Zhang, Y. Y.  Qi,  X. R.  Chen and L. C.  Cai, Phys. Chem. Chem. Phys. \textbf{18},  30061 (2016).



\bibitem{l5}L. M.  Sandonas,D. Teich, R. Gutierrez, T. Lorenz, A. Pecchia, G. Seifert  and G. Cuniberti, J. Phys. Chem. C  \textbf{120}, 18841 (2016).

\bibitem{1}P. Hohenberg and W. Kohn, Phys. Rev. \textbf{136},
B864 (1964); W. Kohn and L. J. Sham, Phys. Rev. \textbf{140},
A1133 (1965).

\bibitem{2}P. Blaha, K. Schwarz, G. K. H. Madsen, D. Kvasnicka
 and J. Luitz, WIEN2k, an Augmented Plane Wave
+ Local Orbitals Program for Calculating Crystal Properties
(Karlheinz Schwarz Technische Universit\"at Wien, Austria) 2001,
ISBN 3-9501031-1-2



\bibitem{pbe}J. P. Perdew, K. Burke and M. Ernzerhof, Phys. Rev. Lett. \textbf{77}, 3865 (1996).

\bibitem{10}A. H. MacDonald, W. E. Pickett and D. D. Koelling, J. Phys. C \textbf{13}, 2675 (1980).

\bibitem{11}D. J. Singh and L. Nordstrom, Plane Waves, Pseudopotentials and the LAPW
Method, 2nd Edition (Springer, New York, 2006).

\bibitem{12}J. Kunes, P. Novak, R. Schmid, P. Blaha and
K. Schwarz, Phys. Rev. B \textbf{64}, 153102 (2001).

\bibitem{so}D. D. Koelling, B. N. Harmon, J. Phys. C Solid State Phys.  \textbf{10}, 3107 (1977).



\bibitem{b}G. K. H. Madsen and D. J. Singh, Comput. Phys. Commun. \textbf{175}, 67
(2006).

\bibitem{b1-1}B. L. Huang and M. Kaviany, Phys. Rev. B \textbf{77}, 125209 (2008).

\bibitem{b2}L. Q. Xu, Y. P. Zheng and J. C. Zheng, Phys. Rev. B \textbf{82}, 195102 (2010).

\bibitem{b3}J. J. Pulikkotil, D. J. Singh, S. Auluck, M. Saravanan, D. K. Misra, A. Dhar and R. C. Budhani,
Phys. Rev. B \textbf{86}, 155204 (2012).

\bibitem{pv1} G. Kresse, J. Non-Cryst. Solids \textbf{193}, 222 (1995).

\bibitem{pv2} G. Kresse and J. Furthm$\ddot{u}$ller, Comput. Mater. Sci. 6, \textbf{15} (1996).

\bibitem{pv3} G. Kresse and D. Joubert, Phys. Rev. B \textbf{59}, 1758 (1999).

\bibitem{pv4}A. Togo, L. Chaput and I. Tanaka, Phys. Rev. B \textbf{91}, 094306 (2015).

\bibitem{2dl}X. F. Wu, V. Varshney et al., Chem. Phys. Lett. \textbf{669}, 233 (2017).

\bibitem{tt9} T. J. Scheidemantel, C. Ambrosch-Draxl, T. Thonhauser, J. V. Badding and
J. O. Sofo, Phys. Rev. B \textbf{68}, 125210   (2003).


\bibitem{tt10} G. K. H. Madsen, J. Am. Chem. Soc. \textbf{128}, 12140 (2006).

\bibitem{tt11} X. Gao, K. Uehara, D. Klug, S. Patchkovskii, J. Tse and T. Tritt, Phys.
Rev. B \textbf{72}, 125202 (2005).


\bibitem{p1}H. Zabel, J. Phys.: Condens. Matter  \textbf{13}, 7679 (2001).

\bibitem{p2} A. H. Castro Neto, F. Guinea, N. M. R. Peres, K. S. Novoselov and A. K. Geim, Rev. Mod.
Phys.  \textbf{81}, 109 (2009).


\bibitem{w3}K. Kutorasinski, B. Wiendlocha, J. Tobola and S. Kaprzyk,
Phys. Rev. B \textbf{89}, 115205 (2014).

\bibitem{w4}S. D. Guo and J. L. Wang, RSC Adv. \textbf{6}, 31272 (2016).

\bibitem{w5}S. D. Guo, J. Alloy. Compd. \textbf{663}, 128 (2016).

\bibitem{e4-1}H. Y. Lv,   W. J. Lu,   D. F. Shao,  H. Y. Lub and   Y. P. Sun, J. Mater. Chem. C \textbf{4}, 4538 (2016).


\bibitem{gsd}S. D. Guo, arXiv:1611.04119 (2016).

\bibitem{b1}C. Kamal and Motohiko Ezawa, Phys. Rev. B \textbf{91}, 085423 (2015).
\end{references}
\end{document}